\documentclass{svproc}
\usepackage{graphicx}
\usepackage{listings}
\usepackage{tabularx}
\usepackage[labelfont=bf]{caption}
\usepackage{subcaption}
\usepackage{url}

\usepackage[leftcaption]{sidecap} 
\sidecaptionvpos{t}{figure}

\usepackage{enumitem}
\setlist[itemize,1]{label=\textbullet}

\newcolumntype{Y}{>{\centering\arraybackslash}X}

\begin{document}

\title{Estimation of Average Annual Daily Bicycle Count Using Bike-Share GPS Data and Bike Counter Data for an Urban Active Transportation Network}

\titlerunning{Estimation of Bicycle Count using Bike-Share GPS Data}  

\author{Marzi Rafieenia \and Liza Wood \and Mohsen Zardadi \and \\ Scott Fazackerley \and Ramon Lawrence}

\authorrunning{Marzi Rafieenia et al.} 

\institute{Department of Computer Science, University of British Columbia, \\ Kelowna, BC, Canada, V1V 2Z3\\
\email{marzi.rafieenia@ubc.alumni.ca, liza.wood@ubc.alumni.ca, mohsen.zardadi@ubc.alumni.ca, \\scott.fazackerley@alumni.ubc.ca, ramon.lawrence@ubc.ca}}

\maketitle  

\begin{abstract}
In 2018, the City of Kelowna entered into a license agreement with Dropbike to operate a dockless bike-share pilot in and around the downtown core. The bikes were tracked by the user’s cell phone GPS through the Dropbike app. The City’s Active Transportation team recognized that this GPS data could help understand the routes used by cyclists which would then inform decision-making for infrastructure improvements. Using OSMnx and NetworkX, the map of Kelowna was converted into a graph network to map inaccurate, infrequent GPS points to the nearest street intersection, calculate the potential paths taken by cyclists and count the number of trips by street segment though the comparison of different path-finding models. Combined with the data from four counters around downtown, a mixed effects statistical model and a least squares optimization were used to estimate a relationship between the different traffic patterns of the bike-share and counter data. Using this relationship based on sparse data input from physical counting stations and bike share data, estimations and visualizations of the annual daily bicycle volume in downtown Kelowna were produced. The analysis, modelling and visualization helped to better understand how the bike network was being used in the urban center, including non-traditional routes such as laneways and highway crossings.
\end{abstract}


\section{Introduction}

The City of Kelowna is located in the Okanagan Valley in the interior of British Columbia and is the third largest city in the province. To combat increasing traffic congestion, the City of Kelowna is developing active transportation corridors within the downtown core. The Kelowna 2030 Official Community Plan envisions \emph{urban communities that are compact and walkable} with \emph{walking paths and bicycle routes to key destinations}~\cite{kelowna:2011}.  The \emph{Kelowna On the Move: Pedestrian and Bicycle Master Plan} was developed to help identify infrastructure, planning and policy requirements that promote and facilitate walking and cycling and realize an Active Transportation Vision~\cite{kelowna:2016}.  

The goal is to improve safety, connectivity, and accessibility by:

\begin{itemize}
    \item Improving the quality and attractiveness of pedestrian and cycling facilities by establishing a low-stress Primary Network for users of all ages and abilities
    \item Reducing conflicts due to truck, transit, and bicycle network overlaps
    \item Enhancing connectivity and continuity with new routes through gap areas
    \item Adding cycling connectivity through high speed, high vehicle traffic volume areas with new connections and direct routes
\end{itemize}

A primary goal is to increase year-round walking and cycling such that  within 20 years, 25\% of all trips less than five kilometers in length are made by walking and cycling~\cite{kelowna:2016}.  Kelowna currently has over 70 km of off-road pathways and 280 km of bike lanes available~\cite{kelowna:2019}.  According to the Pedestrian and Bicycle Master Plan, future infrastructure projects are currently prioritized based on geographic area, closing gaps between existing infrastructure, connectivity to transit, primary network routes and school connectivity. 

To help the city better understand pedestrian and bicycle activity in the  downtown core, a number of traffic counters have been installed.  The challenge is that the counter data only relates temporal traffic flow at a specific point in the transportation network but provides no insight into how users are moving once they enter the core.  To understand the potential movement of cyclists within the network, accurate position data is required.  With the recent interest in bike-share programs such as Dropbike and the adoption of cyclist apps such as Strava~\cite{Lee:2021}, cities now have access to data that can be used to answers questions about cycling behaviours.   While raw GPS data provide some insights into cycling patterns, challenges exist with route determination due to GPS accuracy and sample rates.  Data generated from bike-share programs in the region capture only a fraction of true bike traffic as well as potentially focus on tourism and not general commuting patterns across the cycling network~\cite{Dadashova:2020}.  

In this work, techniques for cleaning and analyzing sparse GPS data are presented and fit to feasible cycling networks.  Data is upscaled using in-network bicycle counter data to develop feasible Average Annual Daily Bicycle counts for segments within the cycling network.  The major steps are: 
\begin{enumerate}
    \item Cleaning and validation of bike-share generated GPS data 
    \item Modelling of sparse GPS data to the City of Kelowna's downtown active transportation network to produce a series of feasible path models for evaluation
    \item Upscaling of counter generated data with feasible path models for evaluation
    \item Computation of estimated Average Annual Daily Bicycle traffic for all segments in the network evaluation area based on preferred path-finding model
\end{enumerate}

The contributions of this work are:

\begin{itemize}
    \item Techniques for cleaning and validating bike-share GPS data within a cycling network based on sparse GPS inputs
    
    \item Comparison of bike-share based network utilization models based on up-scaled counter data
    
    \item Estimation and understanding of Average Annual Daily Bicycling volumes for segments within the downtown core of the City of Kelowna
    
    \item An understanding of how cyclists utilize lane ways and shortcuts within the network
    
\end{itemize}
The paper outline is as follows: Section~\ref{sec:bg} overviews previous approaches and discusses bike-share data. The study area, dataset and data cleaning are discussed in Section~\ref{sec:dataset}. Discussion of models and analysis are presented in Section~\ref{sec:model}. The paper closes with future work and conclusions.

\section{Background}\label{sec:bg}

Cycling in urban centers increases with improved cycling routes~\cite{BARBOUR2019253} and has demonstrated significant time sharing over other commuting means such as taxis.  Bike sharing programs~\cite{KOU2020104534} have a positive contribution on the well being of individuals by increasing physical activities, reducing traffic congestion and improving air quality~\cite{BARBOUR2019253,BOPP:201821,grimes:2020,KOU2020104534,MEDARDDECHARDON2015260}. Analysis has  been expanded to better understand how bike-shares can contribute to the well-being of specific demographic clusters in urban settings~\cite{grimes:2020}.

Bike-share programs are gaining in popularity worldwide as they provide bicycle access to individuals who may not have access to a bicycle. Bike-share systems are categorized as either station-based where a user is required to access and return a bicycle to one of many fixed stations or dockless where the bicycle can be picked up or dropped off at any location~\cite{Wenwen:2020}.  Users access the bike-share through a mobile app that also performs location tracking during the ride.  

Previous works have focused on understanding demand and placement of bicycles in bike-share models, and anticipating how and where bikes should be placed to satisfy demand~\cite{MEDARDDECHARDON2015260,Deep:2020}.  Studies have examined using a combination of bike share GPS data, GPS data from a city app and/or counter data in San Francisco~\cite{proulxbicycle:2017} and Montreal~\cite{STRAUSS:2017155}.  A number of works have also considered data provided by Strava Metro\footnote{https://www.strava.com}, a popular fitness tracking application, from sites in the United States and Europe to better understand active transportation corridor utilization and impacts from weather~\cite{Dadashova:2020,Lee:2021,LIN:2020334}. Works primarily focus on route analysis, weather impacts and infrastructure evaluation and rely on high-frequency data providing detailed insight into cycling behaviours as well as determining the \textbf{Average Annual Daily Bicycle traffic} (AADB) of  street segments in a cycling network.  Equivalent to the fundamental traffic engineering metric of the \textbf{Annual Average Daily traffic} (AADT), the AADB is used to represent the average bicycle traffic  in a given network segment over a year and is commonly used for planning activities~\cite{Nordback:2013,Esawey:2013}.  Data often is sparse and measured over a short duration requiring methods for modelling when used to estimate AADB~\cite{Esawey:2014}.  Analysis of data can  provide new insights into the spatio-temporal dynamics of public bicycle usage~\cite{CORCORAN2014292}.

Challenges exist with  bike-share and app-based data as the data often describes a specific group that is using bike shares, but provides little insight into the general biking population.   Additional challenges exist with the nature of GPS data as it relates to positioning and sampling.  Position sampling may be infrequent and impacted by a cyclist's mobile device leading to a scenario where a cyclist has the opportunity to take multiple paths between the sample points. Further challenges exist due to position inaccuracy.  In ideal conditions, cell phone GPS has accuracy of 4.9m, but that worsens near buildings, bridges and trees~\cite{gps}. Municipalities utilize counters placed in key areas within the transportation network to gain an understanding of temporal traffic patterns. While the counters provide  information about traffic at a specific point in the transportation network, they do not provide insight into how traffic flows between counters and within segments of the transportation network.  

Graphs are used to visualize and study the connections between items, and are represented by nodes connected by edges. Graphs can represent transportation networks~\cite{Majeed:2020} and street maps~\cite{STEPANCHUK:2016276}.  With street maps, the nodes represent intersections while the edges represent streets, lanes, or bike paths. 

Numerous packages are available to represent maps as graphs for analysis. OSMnx~\cite{boeing:2017} and NetworkX~\cite{networkx:2020} are Python packages that provide tools to create and analyze maps as graphs. OSMnx downloads street networks from OpenStreetMap\footnote{https://www.openstreetmap.org/}, which is supported by a community of developers. OSMnx is for the construction, projection, visualization, and analysis of complex street networks in Python with NetworkX~\cite{boeing:2017}.

Graph edges can be given a numerical weight to represent the distance or cost between nodes. The \verb|shortest_path| algorithm calculates the path between two nodes with the minimum weight producing the shortest total path length. As cyclists do not always travel the shortest distance, numerous potential paths can be determined based on different weighting strategies.  

Prediction and analysis of cycling behaviors can be conducted by using counter and GPS bike data modelled in a cycling network.  Numerous challenges exist with fitting GPS data to a feasible network.  For systems utilizing both GPS and counter data, significant work is required to  determine a statistically significant relationship between the bike-share and counter data. 

\section{Study Area, Dataset and Data Cleaning}\label{sec:dataset}

\begin{figure}[!t]
\begin{subfigure}[t]{0.32\columnwidth}
   	\centering
	\includegraphics[width=\columnwidth]{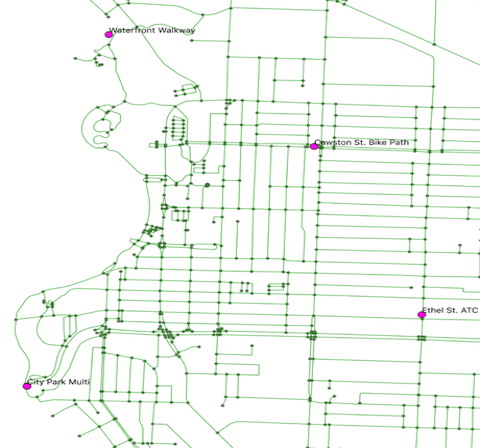}
	\caption{Kelowna transportation network.}
	\label{fig:kelowna_network}
\end{subfigure}
\captionsetup[subfigure]{justification=centering}
\begin{subfigure}[t]{0.32\columnwidth}
    \centering
	\includegraphics[width=\columnwidth]{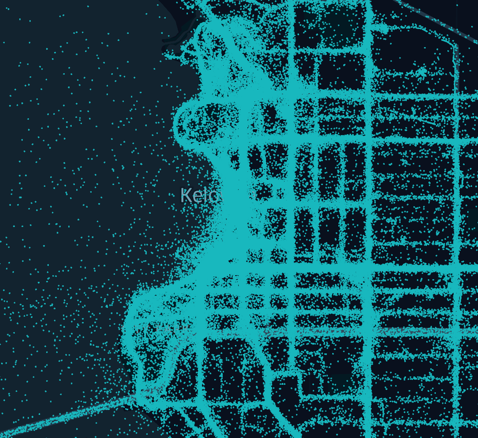}
	\caption{GPS datapoints}
	\label{img:gps_1}
\end{subfigure}
\begin{subfigure}[t]{0.32\columnwidth}
    \centering
	\includegraphics[width=\columnwidth]{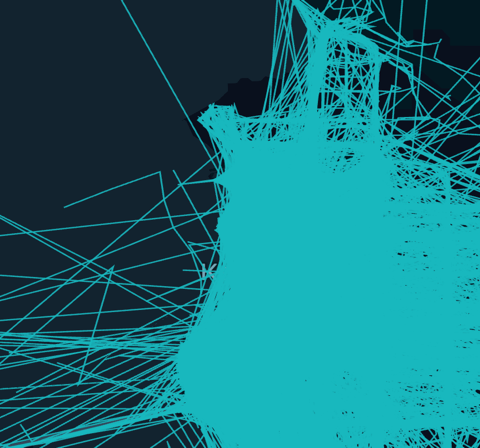}
	\caption{Lines connecting bike-share GPS points}
	\label{img:gps_2}
\end{subfigure}
\caption{Study area and bike-share datapoints in the study area of the downtown core of Kelowna, BC, Canada with counting stations in pink}\label{fig:gps}
\end{figure}

In February 2018, the City of Kelowna entered into an agreement with Dropbike\footnote{https://dropmobility.com} to operate a bike-share pilot in and around the downtown core (Figure~\ref{fig:kelowna_network}) bounded by the City Park Multi (south-west corner), Waterfront Walkway (north-west corner), Cawston Street Corridor (north-east corner), and Ethel Street ATC (south-east corner).   Dropbike is a dockless system, so users were not confined to starting and stopping their trips at specific locations. The bikes were tracked by a user’s cell phone GPS with the Dropbike app.  The City recognized that the pilot would provide data to understand the routes used by cyclists and inform decision-making for infrastructure improvements.

The pilot ran from 11 June to 11 November 2018. During the first three months of the pilot, there were more than 33,000 (trips) from 9,000 unique users on 331 bikes.  Initial analysis indicated that approximately a third of the total trips were made by 600 frequent users. From the trial period, the City of Kelowna has Dropbike GPS data for each uniquely identifiable trip.  Visualizations (Figure~\ref{img:gps_1}) plot the raw GPS data points for all rides during the trial, providing a sense of where bike-share riders have cycled, along with relative density of bike-share traffic. When the points are connected to visualize the individual routes (Figure~\ref{img:gps_2}), the routes are indistinguishable. A more quantified analysis was needed to inform decision-making for infrastructure improvements.  In addition, bike-share data only represents a portion of all the bicycle traffic in Kelowna.

To understand pedestrian and bicycle activity around Kelowna, 38 traffic counters are installed in key locations along transportation corridors. Of the counters, 19 counters have the ability to monitor bicycle traffic and can distinguish the direction of traffic. These counters collect real-time data and upload the data to the city via cell uplink once per day. Since they have started collecting data at their respective locations, the city-wide counters have counted 6,496,174 bicycle trips to date over a period of seven years.  Analysis was restricted to traffic in the bounded urban core with four counters at the main access points.

\subsection{Challenges with Bike-share GPS Data}

During the first 13 weeks of the pilot nearly 9000 trips of data were gathered. This is an average of 99 cyclists per day, a small portion of which pass one of the four counters. During this same period, the four counters counted 232,835 cyclists, an average of 640 cyclists per counter per day. Bike-share trips are small compared to the total volume of bicycle traffic. 

\begin{figure}[!t]
\centering
\begin{minipage}{.48\textwidth}
  \centering
  	\includegraphics[width=\columnwidth]{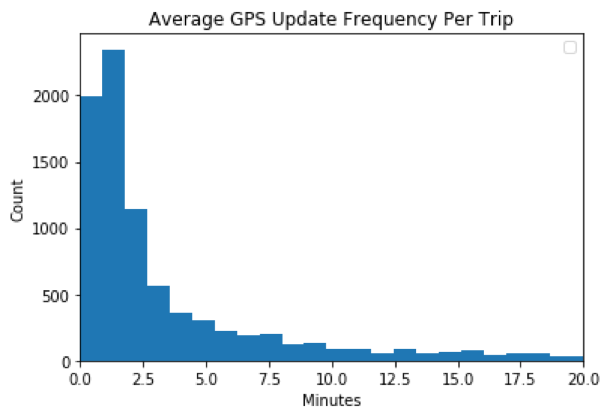}
	\caption{Distribution of average GPS update frequency from bike-share data}
	\label{fig:gps_update_time}
\end{minipage}%
\hfill
\begin{minipage}{.48\textwidth}
  \centering
 \includegraphics[width=\columnwidth]{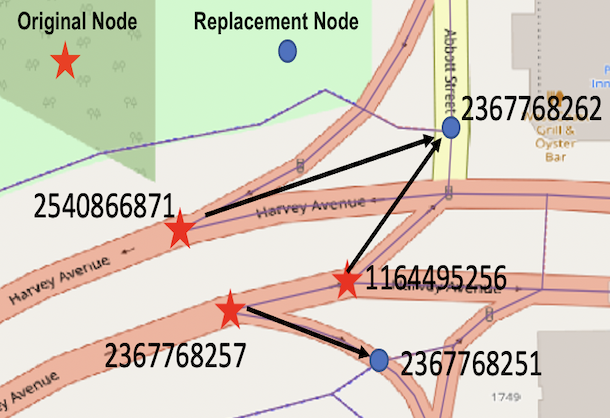}
	\caption{Remapping of problem nodes at Highway 97 and Abbott Street}
	\label{fig:remap}
\end{minipage}
\end{figure}

In the dataset generated from bike-share data, numerous GPS points are far away from the nearest possible bicycle transportation corridor (Figure~\ref{img:gps_1}), including some data in the middle of Okanagan Lake.  Bicycle GPS data points need to be assigned to feasible bicycle paths and a feasible route determined between each of the points per trip.  The frequency of the GPS updates for most of the trips is low and not consistent (Figure~\ref{fig:gps_update_time}).  

The median update frequency is 2 minutes 18 seconds, during which an average cyclist could travel up to 400 meters. This creates a scenario where a cyclist could take one of multiple feasible routes between two GPS data points.  

In the process of performing their own analysis of the bike-share data, the City of Kelowna identified which bike-share trips and GPS points needed to be cleaned out of the data set based on realistic minimum and maximum trip lengths, realistic urban cycling speeds and trip times in addition to segment specific rules. While the trips and points violating these criteria were identified, they were still in the data set, representing 22153 trips. Outliers that did not meet the identified criteria were filtered out, resulting in 8853 valid trips based on 97110 GPS points. 

Using NetworkX, it was possible to find the path between any two nodes in a graph. The first step in the analysis used the \verb|get_nearest_node| algorithm in OSMnx to calculate the nearest valid transportation network node to each GPS point and the corresponding distance. The calculations needed to find the nearest node for each GPS point utilized Google’s Colaboratory\footnote{https://research.google.com/colaboratory}, which allows Python code to be executed in a web browser.  It is well suited for machine learning and data analysis problems, reducing computation time with cloud computing. 
With the nearest node, each GPS point was represented by its nearest intersection on the map. GPS points in Okanagan Lake or outside of Kelowna were excluded from the analysis. The distance to the nearest node was used to filter out GPS points that were greater than 150m from the calculated nearest node. With those points removed, any trips with less than three points remaining were removed resulting in 8,815 trips and 95,905 GPS points.

The \verb|shortest_path| algorithm in NetworkX was then used to calculate the path between nodes. There were 72 trips for which the algorithm could not find a path of which 59 passed through Highway 97 and Abbott St. intersection.  This is an intersection that crosses a major highway route with a complex crossing pattern.  Challenges existed with GPS inaccuracy in the area and possible feasible solutions as well as how cyclists crossed the intersection.  To resolve this issue, problem nodes were mapped to the nodes of the path the cyclists are expected to take to cross that intersection (Figure~\ref{fig:remap}). In the remaining problem trips nodes did not connect back to a road or cycleway due to the unconstrained cycling options in these areas. They were fixed using the same methodology of remapping to an expected feasible solution. 

\section{Model Assessment, Analysis and Results}\label{sec:model}
\begin{table}[!t]
\caption{Path segment weighting strategies}
\label{tab:path_weights}
\begin{tabularx}\columnwidth{l|Y}
\hline
\multicolumn{1}{c|}{\textbf{Weighting Method}} & \multicolumn{1}{c}{\textbf{Description}}                                               \\ \hline
Length                                         & meters of street/path segments in a route                                              \\ \hline
Path Preference (PP)~\cite{winters:2010}                                & a numerical value given to each street/path category based on cyclist path preferences \\ \hline
Simplified Path Preference (SPP)                    & numerical preference given to cycleways only                                           \\ \hline
Weighted Length (WL)                              & length of the street multiplied by the simplified preference value for cycleways       \\ \hline
Closeness Centrality (CC)~\cite{newman:2008mathematics,ALATTAR:2021100301}                          & streets that have a higher degree of closeness centrality are preferred                \\ \hline
Unbiased (UB)                                       & all street weights are set to one                                                      \\ \hline
Corner-weighted (CW)        & weights adjusted for specific cycleways to match the split of bike-share volumes at counter locations with the same split for the counter data \\ \hline
Corner-weighted Length (CWL) & the length of the street multiplied by the weights from the Corner-weighted values for an edge                                                 \\ \hline
\end{tabularx}
\end{table}

Seven path-finding models were examined as described in Table~\ref{tab:path_weights}. Each of the path-finding models were assessed and the best feasible graph was constructed for the cycling solutions based on the cleaned bike-share data as a representation of traffic throughout the network. 
Attempts were made to fit possible models to upscale and model possible bicycle traffic volumes and paths based on overall counter data.  Factors considered when determining the suitability of each model include feasible shortest paths with valid speed constraints.  Each of the path-finding models were evaluated based on:

\begin{itemize}
    \item Visualization of counts
    \item Percentage of segments within realistic speed bounds within segments
    \item Split of total bike-share counts at each of the counter locations compared with the split of counter counts
    \item Linear regression of bike-share counts versus the counter data
\end{itemize}

Once the paths for each trip were calculated in each of the path-finding models, the count of how many bike-share trips travelled along each street segment was determined.  The path for each trip was broken down into pairs of nodes where each pair of nodes represents the intersections that bound each street or path segment. Each pair of nodes was summed to get the total number of trips for each segment. The path-finding model that performed the best in these evaluations was then used to determine the AADB for segments in the network.

\begin{table}[!t]
\caption{Mapping of cyclist path preference to weight}
\label{tab:preference_weight_mapping}
\begin{tabularx}\columnwidth{Y|Y||Y|Y}
\hline
\multicolumn{2}{l||}{\textbf{Path Preference}} &
  \multicolumn{2}{l}{\textbf{Assigned Weights for Path-finding}} \\ \hline
\multicolumn{1}{c|}{\textbf{Path Type}} &
  \textbf{\begin{tabular}[c]{@{}c@{}}Preference \\Factor\\(Higher is\\ preferred)\end{tabular}} &
  \multicolumn{1}{c|}{\textbf{Path Type}} &
  \multicolumn{1}{c}{\textbf{\begin{tabular}[c]{@{}c@{}}Weight\\ (Lower is\\preferred)\end{tabular}}} \\ \hline
Paved off-street cycling  &
  0.5 &
  Cycle path + Ethel &
  0.5 \\ \hline
Unpaved off-street cycling &
  0.4 &
  N/A downtown &
   \\ \hline
Residential streets &
  0.1 &
  Residential &
  0.9
   \\ \hline
 &
   &
  Lanes, Unclassified &
  1  \\ \hline
Major streets with bike lanes &
  -0.1 &
  Not used &
   \\ \hline
Major streets with bike symbols &
  -0.2 &
  Secondary/Tertiary &
  1.2 \\ \hline
Major streets with parked cars &
  -0.5 &
  Not used &
   \\ \hline
 &
   &
  Hwy 97 &
  3 \\ \hline
\end{tabularx}
\end{table}

For the path preference based models, Table~\ref{tab:preference_weight_mapping} shows values developed in~\cite{winters:2010} mapped to weights in the \emph{Path Preference} and \emph{Simplified Path Preference} path-finding models.  Laneways and footpaths were considered with neutral weights as they provide viable path options for the cycling network. Highway 97 (major transit route for through-city traffic) was weighted to avoid all trips as no cycling infrastructure exits along this corridor.  

The \emph{Corner-weighted} and \emph{Corner-weighted Length} path-finding models were created to get the same split of bike-share counts at the counter locations. The counters are roughly at the four corners of the map (Figure~\ref{fig:kelowna_network}). 
Similar in approach to the \emph{Simplified Path Preference}, the weightings were iteratively adjusted until the split of the bike-share counts at the counter locations were close to the same as those for the counter data, starting with weightings that represented the split of traffic between the counters. The weightings were increased or decreased until the counts from \verb|shortest_path| algorithm resulted in the bike-share counts having a similar split as the counter data for each location.
 
\subsection{Count Visualization}

Visualizing the counts was conducted using QGIS for each model. Areas with no counts, wide variations in continuous paths, and a mismatch in counts compared with the GPS point density indicated a problem with the model. 

\begin{figure*}[!t]
\captionsetup[subfigure]{justification=centering}
\begin{subfigure}[t]{0.32\textwidth}
	\centering
	\includegraphics[width=\textwidth]{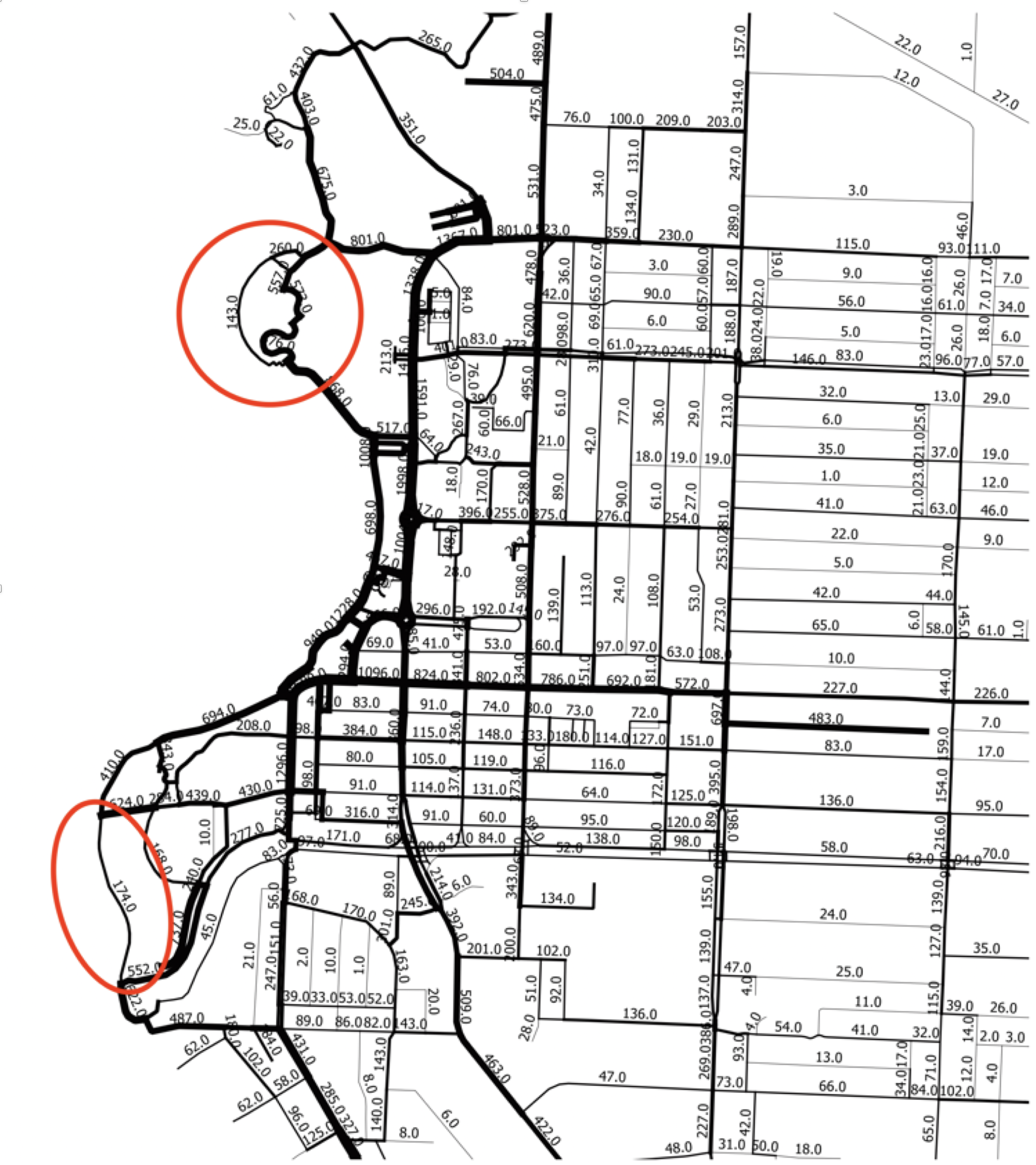}
	\caption{Segment counts based on the shortest length model}
	\label{img:bs_shortest_length_model}
\end{subfigure}
\begin{subfigure}[t]{0.32\textwidth}
\centering
	\includegraphics[width=\textwidth]{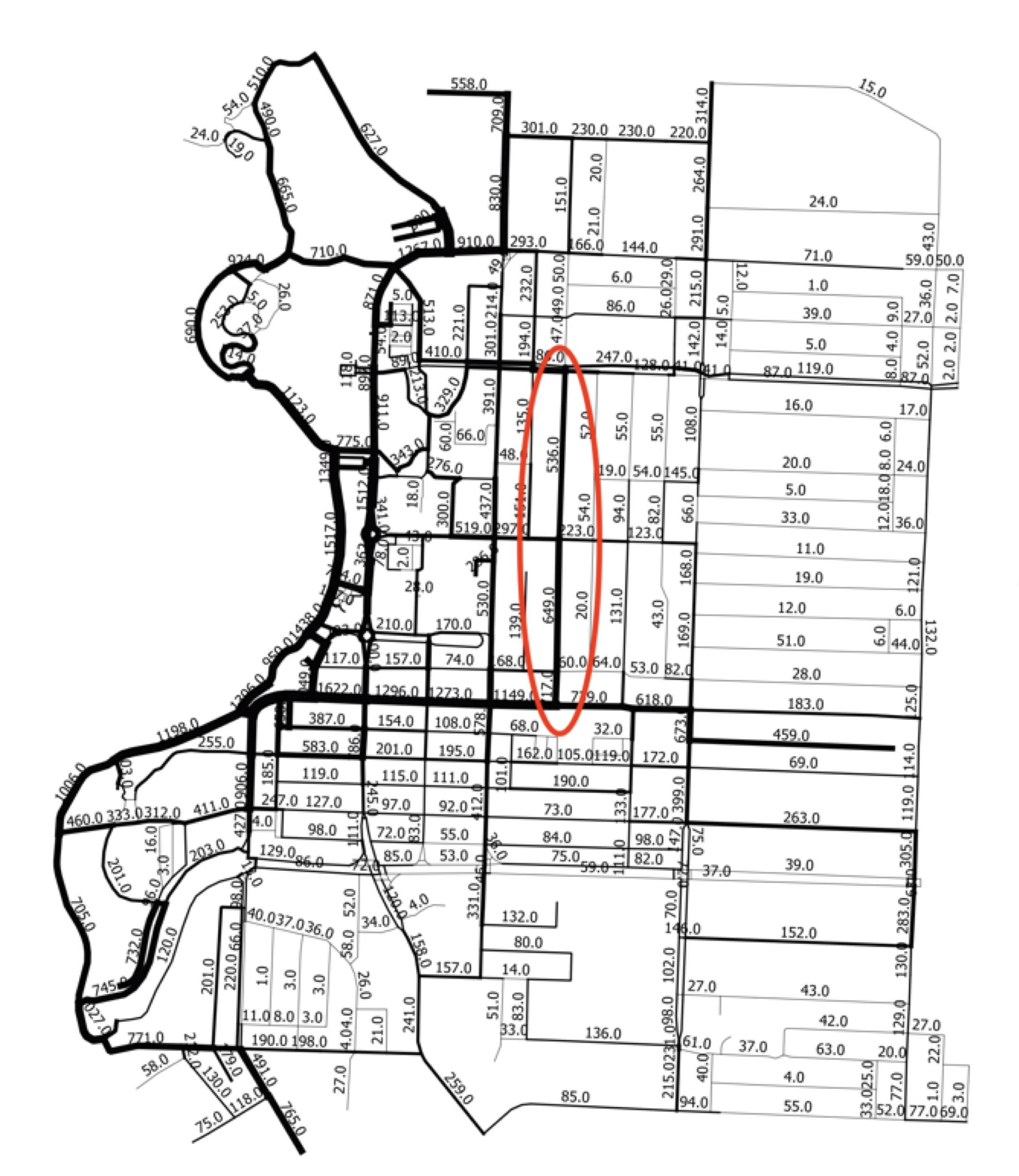}
	\caption{Segment counts based on the closeness centrality model}
	\label{img:bs_CC_model}
\end{subfigure}
\begin{subfigure}[t]{0.32\textwidth}
\centering
	\includegraphics[width=\textwidth]{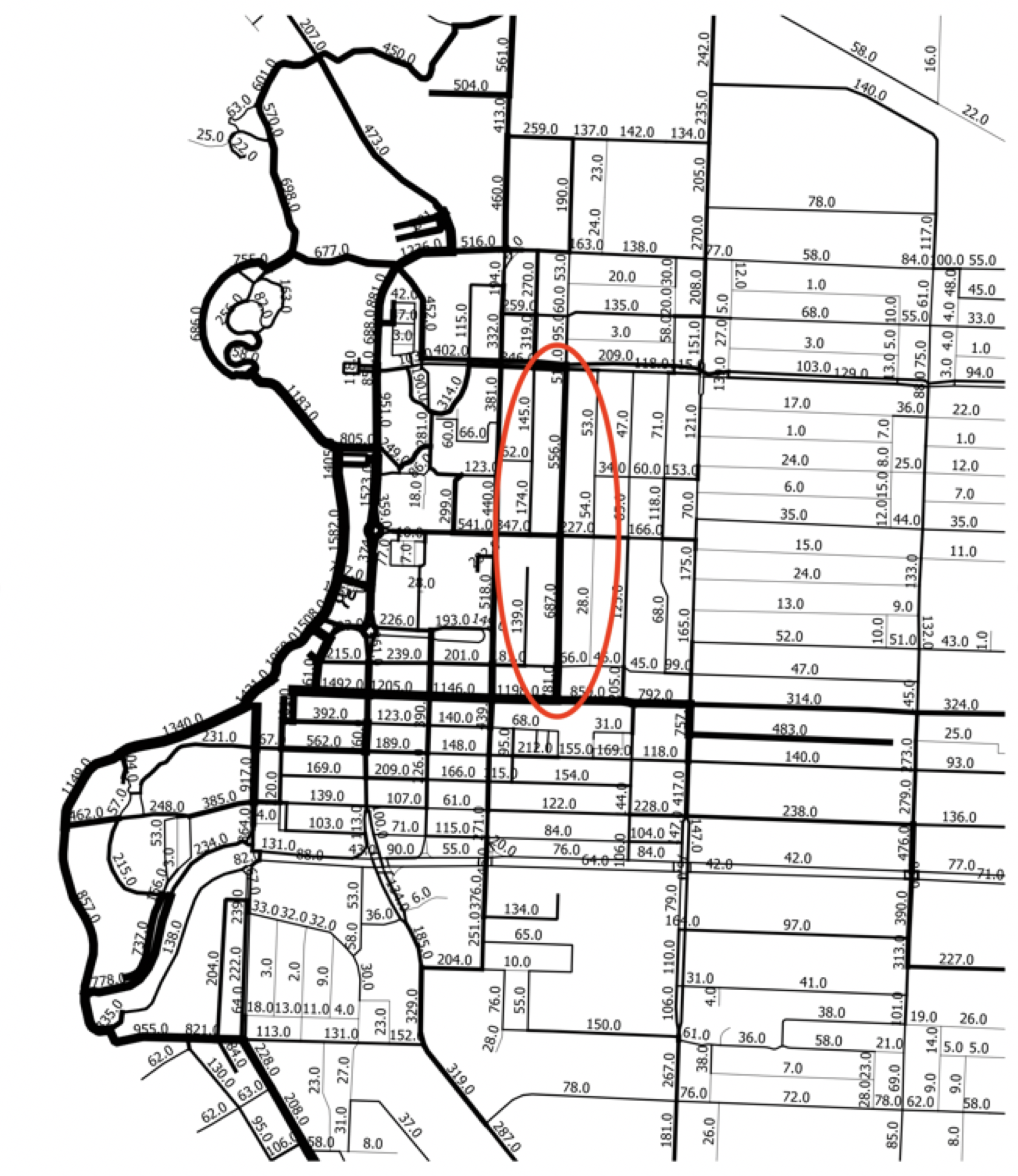}
	\caption{Segment counts based on the unbiased model}
	\label{img:bs_unbiased_model}
\end{subfigure}
\caption{Bike-share counts per segment path with different path-finding models.}
\end{figure*}

Figure~\ref{img:bs_shortest_length_model} presents the counts per system segment based on the \emph{Shortest Length} model.  With this mode, two areas of concern where detected, highlighted by the red circle in the figure. In the top left, more cyclists were counted along the sidewalk by Island Stage in Waterfront Park than along the boardwalk. In the bottom left, there is a noticeably smaller count in the lakeside segment after the City Park counter than in the rest of the segments along that path. The density of GPS points was relatively uniform throughout that area.  

Figure~\ref{img:bs_CC_model} presents counts per system segment based on the \emph{Closeness Centrality} model.  It was observed that St. Paul Street (indicated by a red circle in Figure~\ref{img:bs_CC_model}) has the highest closeness centrality in the model area, and as a result the visualization has an unexpectedly high count.

Figure~\ref{img:bs_unbiased_model} presents counts per system segment based on the \emph{Unbiased} model. St. Paul Street and the surrounding area (circled in red) has unusually high counts along the corridor. These relatively high counts were not reflected in the relative density of GPS points in those areas (Figure~\ref{img:gps_1}).

Additional path-finding models did not present these issues. After reviewing the \emph{Shortest Length}, \emph{Path Preference}, \emph{Closeness Centrality}, \emph{Unbiased}, and \emph{Weighted-Corner} models with the subject-matter experts from the City of Kelowna, three path-finding models based on the visualizations were selected: \emph{Path Preference}, \emph{Weighted-Corner} and \emph{Shortest Length}. 

\subsection{Segment Speed Boundaries}

A key consideration in model determination was the assessment of valid bicycle speeds along a given segment.  As the bike-share data for some trips is sparse, the shortest path connection point resulted in speeds that were unrealistic for cyclists.  The speed required to travel along the path between GPS points was used to measure if the calculated path was feasible. If the speed is too high, it is unlikely that path was travelled. To determine the maximum speed, the speed between each pair of GPS points along the length of the calculated path in the \emph{Shortest Length} model was calculated as a upper boundary, using the timestamp for each GPS point. A distribution of speeds from all segments was computed, removing the outliers being any infeasible speeds. It was observed that the maximum feasible speed is 22.5 km/h, and this value was used to set the maximum for evaluating the other path-finding models. Based on the highest percentage of feasible segments, the \emph{Shortest Length}, \emph{Closeness Centrality} and \emph{Corner-Weighted} were selected as candidate models under this condition. 

\subsection{Split of bike-share Counts based on Counter Location}

\begin{table}[!t]
\caption{Bike-share split between counter locations by path-finding models}\label{tab:counter-split}
\begin{tabularx}\columnwidth{l|Y|Y|Y|Y}
\hline
                            & \textbf{City Park} & \textbf{Waterfront} & \textbf{Cawston} & \textbf{Ethel} \\ \hline
Counter Data Split          & 40.7\%             & 16.5\%              & 28.4\%           & 16.5\%         \\ \hline
\textbf{Path-Finding Model} &                    &                     &                  &                \\ \hline
Shortest Length               & 41.3\%             & 26.0\%              & 22.9\%           & 9.7\%          \\
Path Preference             & 45.5\%             & 31.3\%              & 14.2\%           & 9.0\%          \\
Simplified Path Preference  & 47.0\%             & 25.5\%              & 15.4\%           & 12.1\%         \\
Weighted Length             & 39.8\%             & 23.5\%              & 26.6\%           & 10.1\%         \\
Closeness Centrality        & 53.4\%             & 24.5\%              & 15.8\%           & 6.3\%          \\
Unbiased                    & 51.7\%             & 24.1\%              & 12.8\%           & 11.4\%         \\
Corner Weighted             & 44.7\%             & 17.1\%              & 12.5\%           & 16.6\%         \\
Corner Weighted Length      & 35.4\%             & 36.3\%              & 22.9\%           & 5.4\%          \\ \hline
\end{tabularx}
\end{table}

One of the base assumptions of this project was that the bike-share data is representative of all bicycle traffic during the same period. Therefore, the best model would result in a similar split of bike-share counts at the counter locations as seen in the counter data. The actual split of raw counter data from the four stations is presented in row 1 of Table~\ref{tab:counter-split}.   The remaining rows of Table~\ref{tab:counter-split} present the adjusted split for each path-finding model.  A key question that arose in the analysis of the data is if the bike-share data is representative of all bike traffic. Examination of the model splits noted a lack of correlation.

One issue noted is that the model and counter data is presented as a daily aggregate but in practice, bike traffic volumes at the counter points demonstrate diurnal variation.   To better understand the relationship between total bike-share counts and the raw counter data for each counter location, data was compared on an hour-by-hour basis. 

\begin{figure*}[!t]
\captionsetup[subfigure]{justification=centering}
\begin{minipage}[t]{\textwidth}
\begin{subfigure}[t]{0.245\textwidth}
\centering
\includegraphics[width=\textwidth]{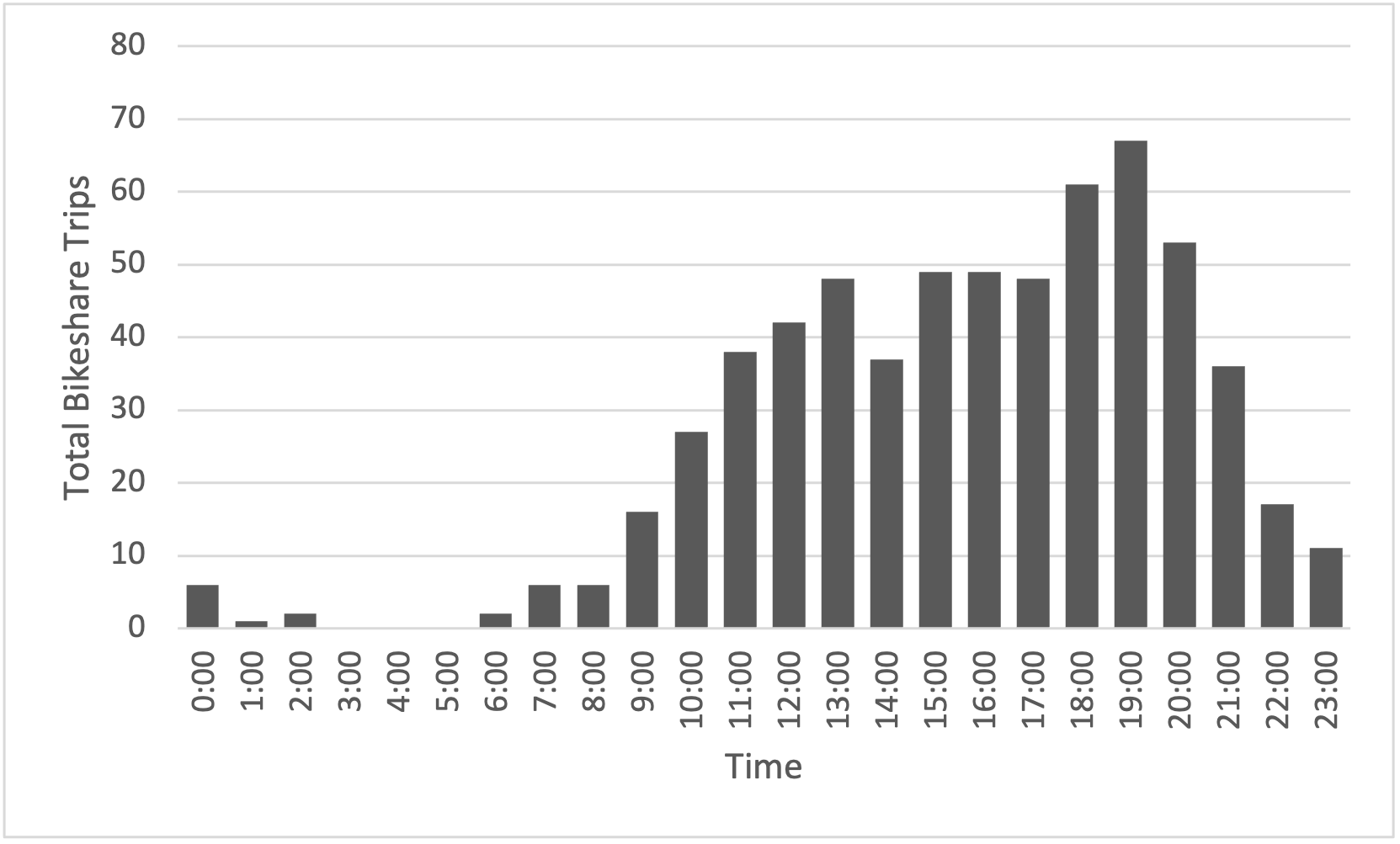}
\caption{City park}
\label{fig:BS_citypark_hourly}
\end{subfigure}
\begin{subfigure}[t]{0.245\textwidth}
\centering
\includegraphics[width=\textwidth]{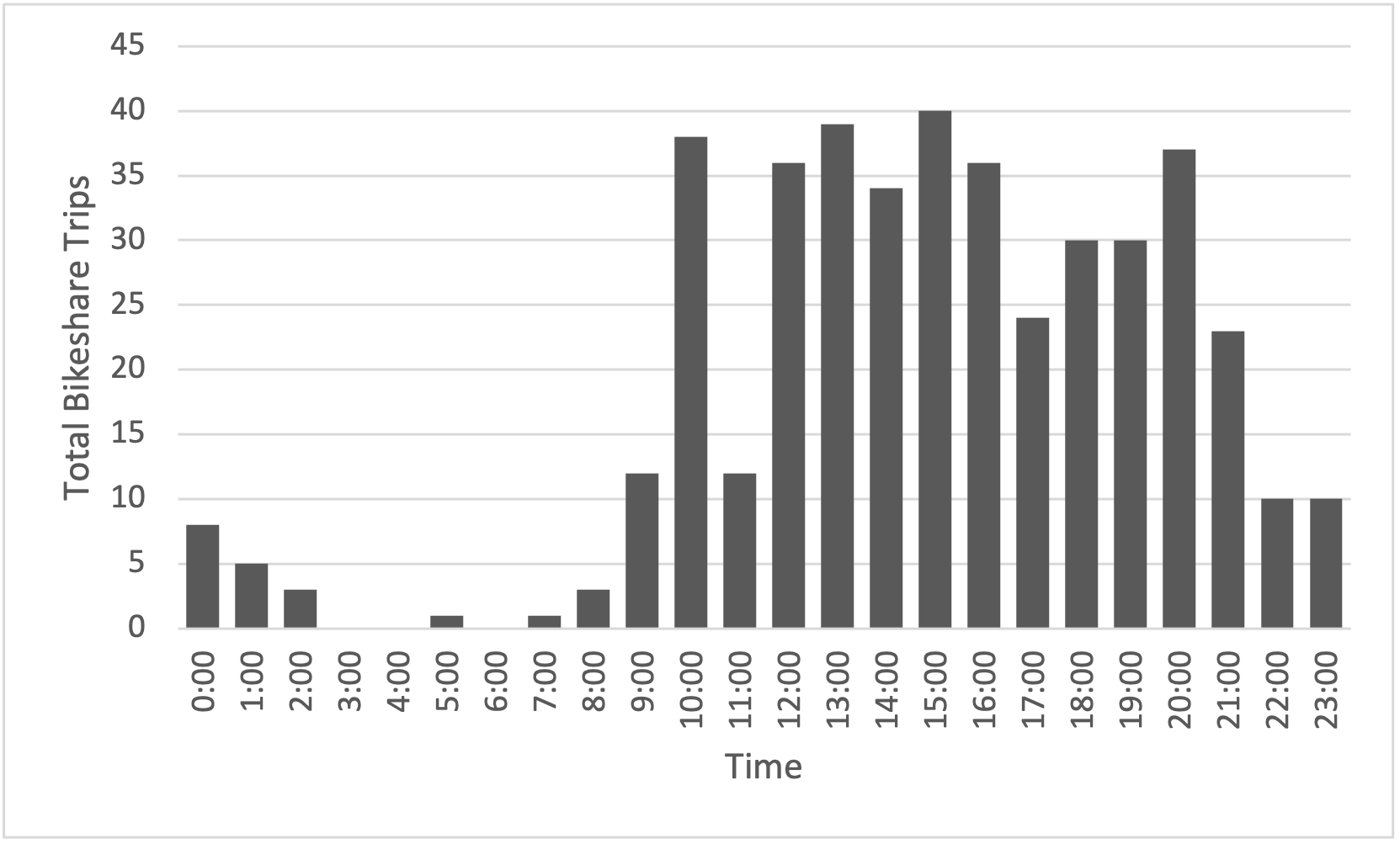}
\caption{Waterfront}
\label{fig:BS_waterfront_hourly}
\end{subfigure}
\begin{subfigure}[t]{0.245\textwidth}
\centering
\includegraphics[width=\textwidth]{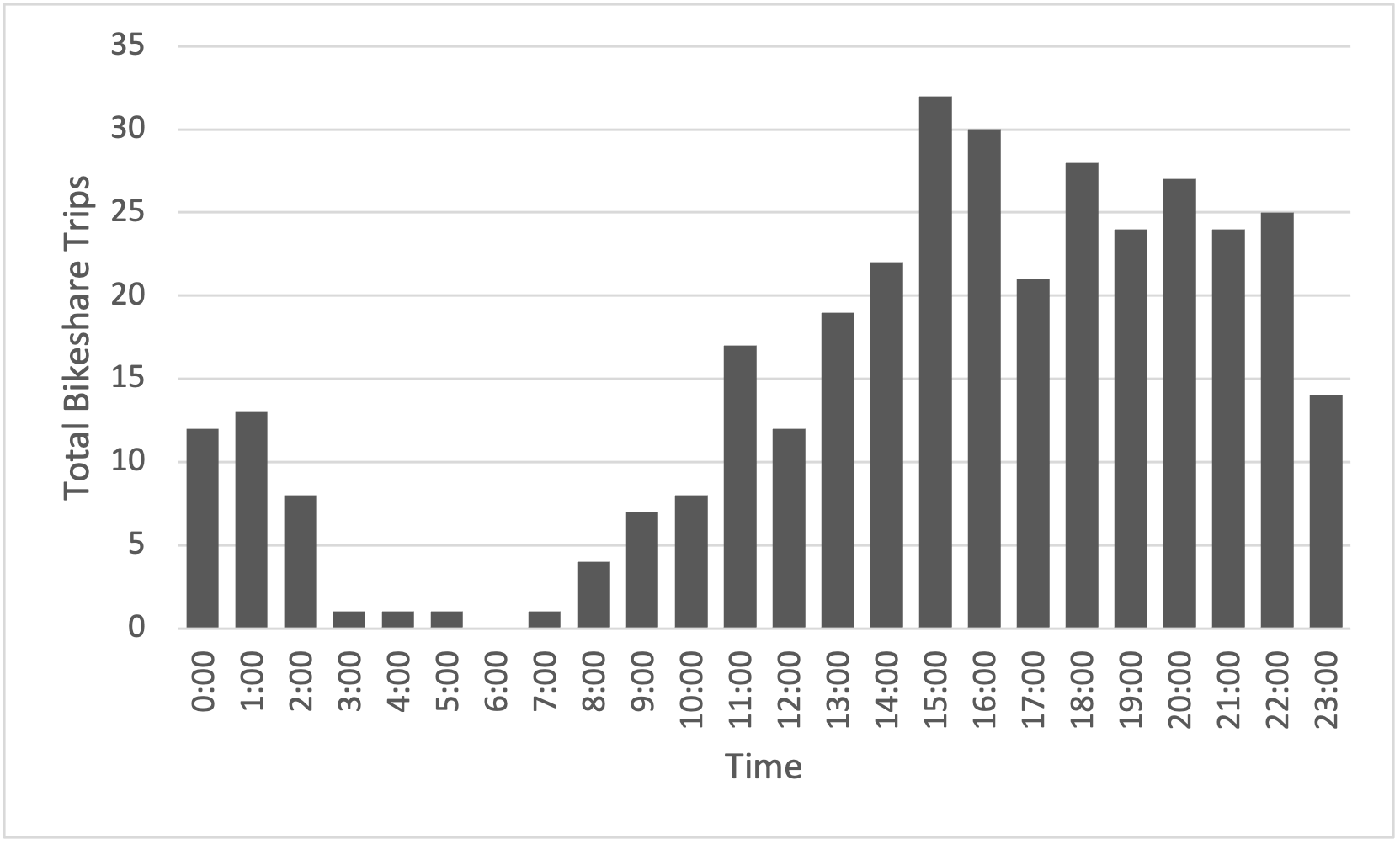}
\caption{Cawston}
\label{fig:BS_cawston_hourly}
\end{subfigure}
\begin{subfigure}[t]{0.245\textwidth}
\centering
\includegraphics[width=\textwidth]{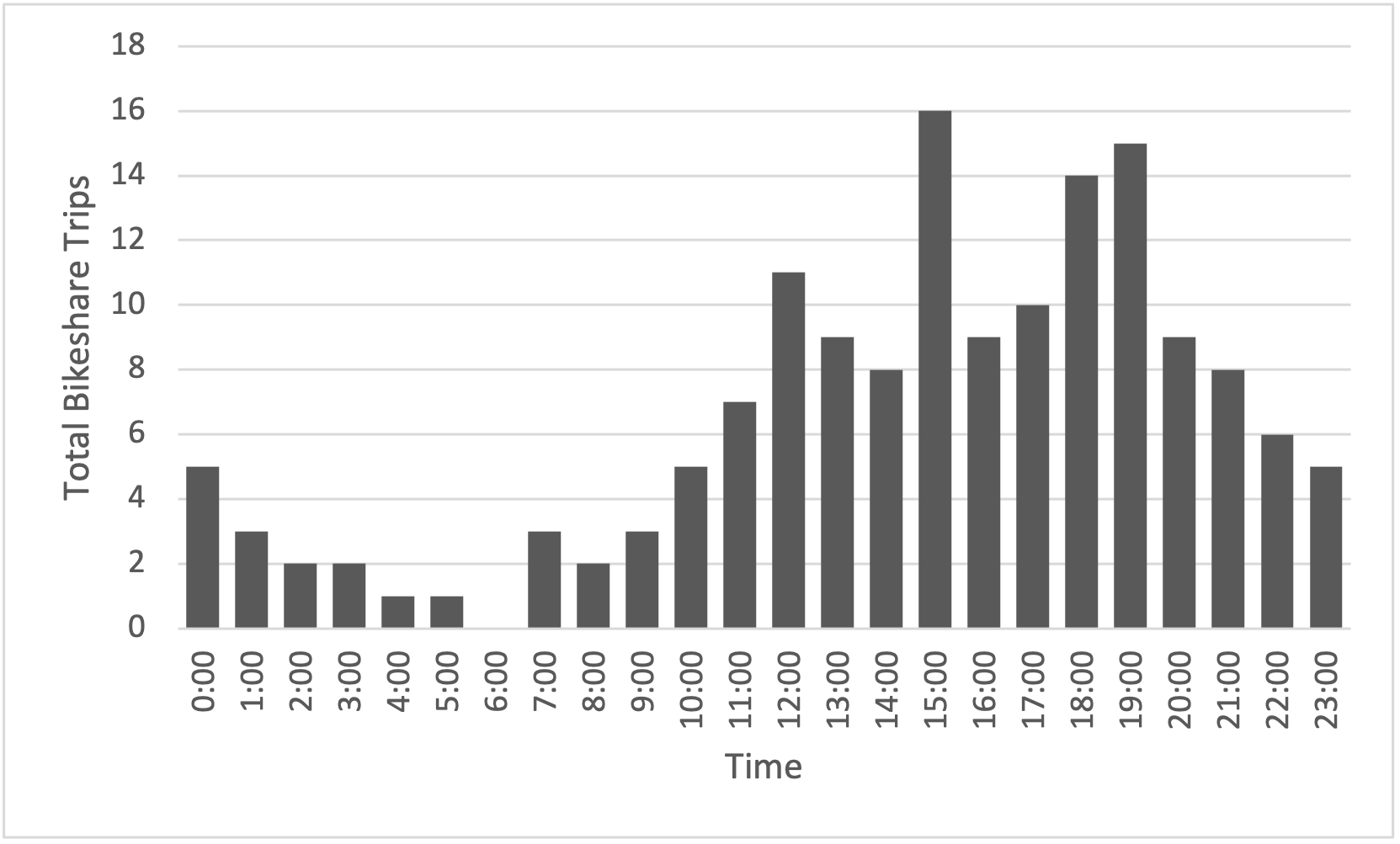}
\caption{Ethel}
\label{fig:BS_ethel_hourly}
\end{subfigure}
\caption{Hourly distribution of bike-share data at data capture points}\label{fig:hourly_bike-share}

\end{minipage}

\vfill

\begin{minipage}[t]{\textwidth}

\captionsetup[subfigure]{justification=centering}
\begin{subfigure}[t]{0.245\textwidth}
\centering
\includegraphics[width=\textwidth]{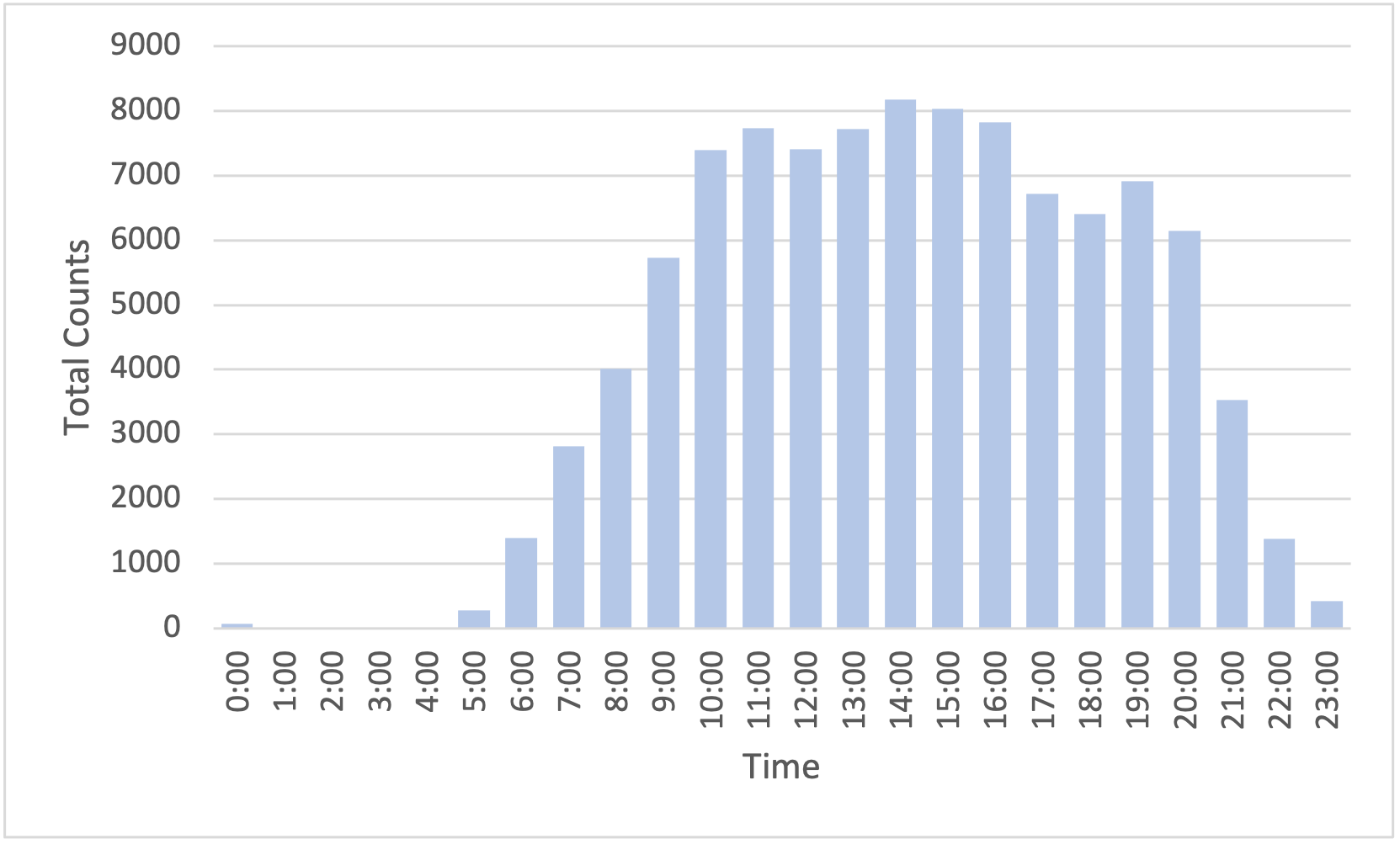}
\caption{City park}
\label{fig:C_citypark_hourly}
\end{subfigure}
\begin{subfigure}[t]{0.245\textwidth}
\centering
\includegraphics[width=\textwidth]{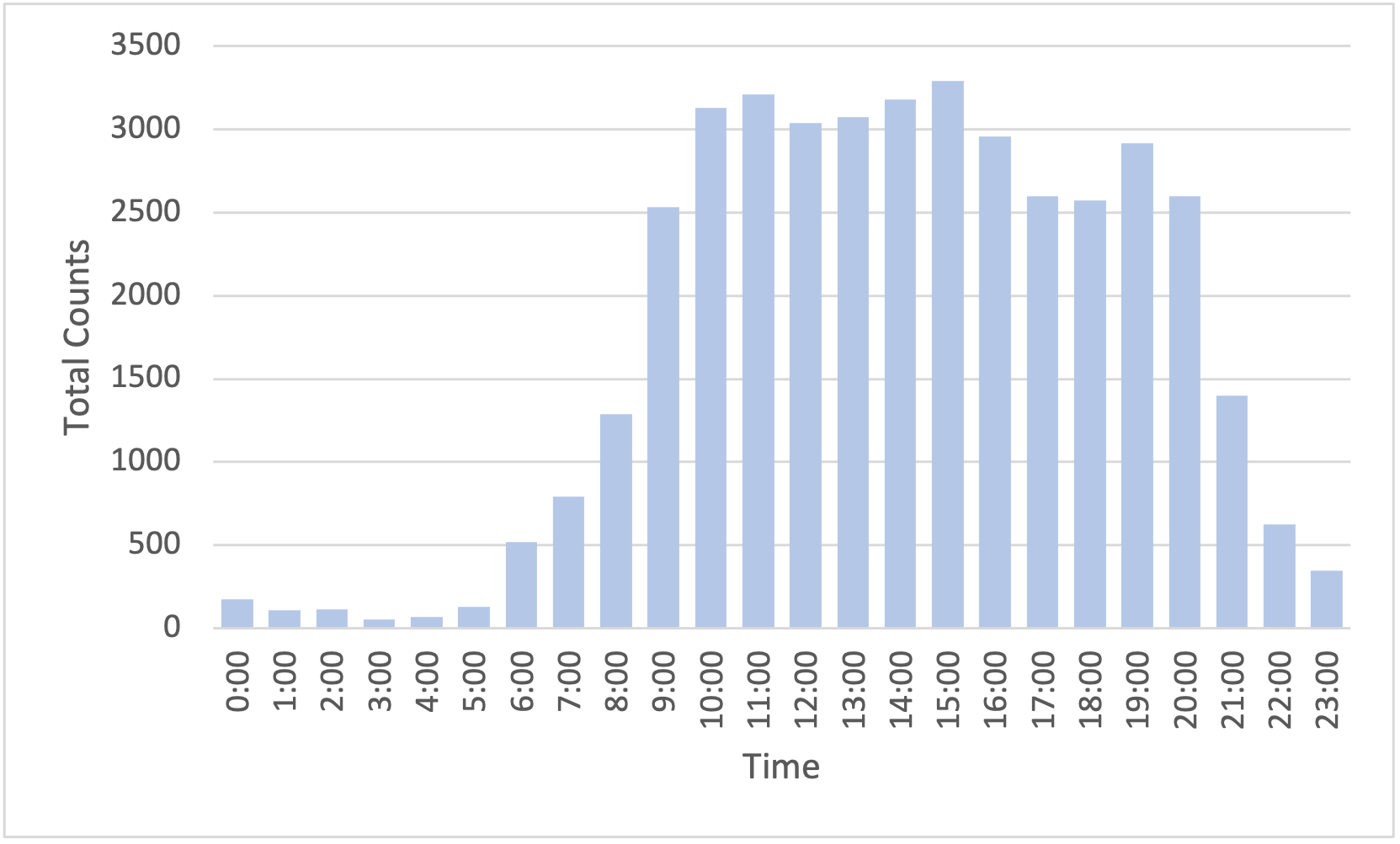}
\caption{Waterfront}
\label{fig:C_waterfront_hourly}
\end{subfigure}
\begin{subfigure}[t]{0.245\textwidth}
\centering
\includegraphics[width=\textwidth]{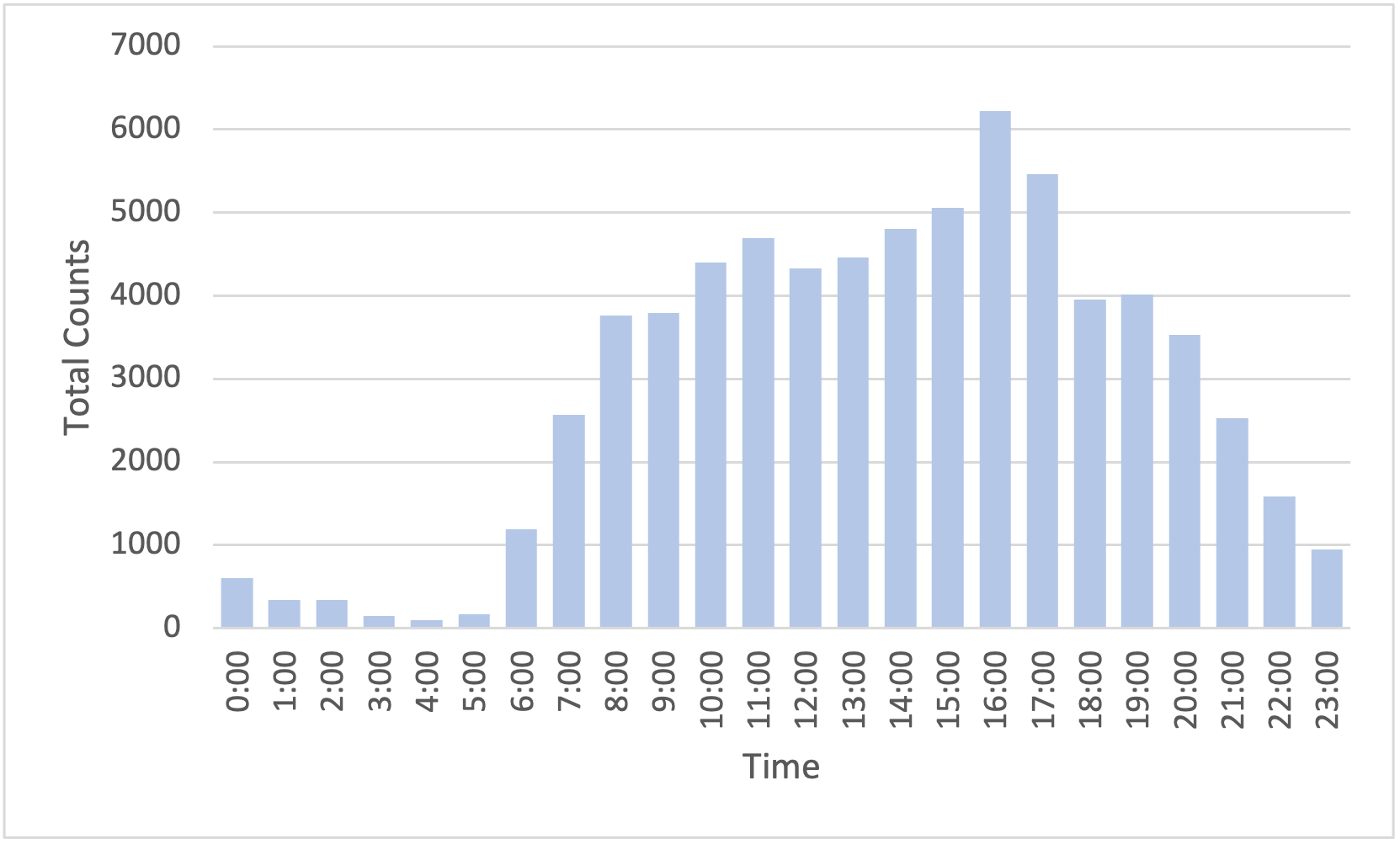}
\caption{Cawtson}
\label{fig:C_cawston_hourly}
\end{subfigure}
\begin{subfigure}[t]{0.245\textwidth}
\centering
\includegraphics[width=\textwidth]{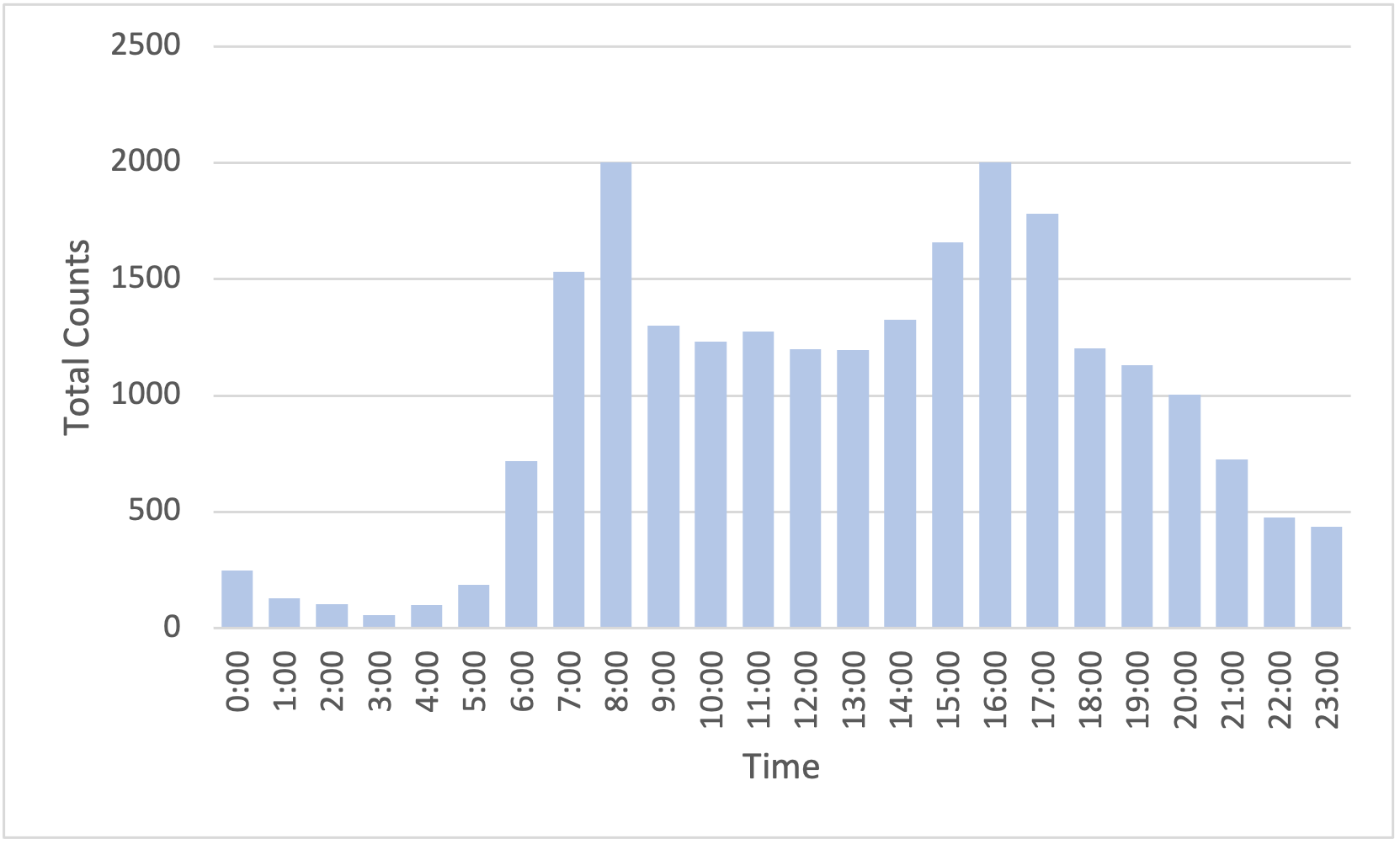}
\caption{Ethel}
\label{fig:C_ethel_hourly}
\end{subfigure}
\caption{Hourly distribution of counter data at data capture points}
\label{fig:hourly_counter}
\end{minipage}

\end{figure*}

Figures~\ref{fig:hourly_bike-share} and \ref{fig:hourly_counter} present the hourly breakdown of count totals for each of the four data capture points. The shape of the distributions  is most similar for the comparison of bike-share and counter values for  City Park (Figures~\ref{fig:BS_citypark_hourly} and \ref{fig:C_citypark_hourly}) and for Waterfront (Figures~\ref{fig:BS_waterfront_hourly} and \ref{fig:C_waterfront_hourly}) locations. Both of those locations are at the boundary to a large recreational corridor in the downtown core and have a high degree of recreational traffic, which is supported by the bike-share data.

When comparing the Cawston (Figures~\ref{fig:BS_cawston_hourly} and \ref{fig:C_cawston_hourly}) and Ethel (Figures~\ref{fig:BS_ethel_hourly} and \ref{fig:C_ethel_hourly}) locations, the shape of the distributions are quite different.  These locations have peaks in the morning and the evening.  This represents a commuter traffic pattern as both capture points are located along a bike commuter pathway leading to the downtown core from surrounding low and medium density housing areas rather than the commercial/tourist area of Waterfront and City Park. The bike-share distributions for these capture points do not have those same peaks as this area has a large recreation focus along a waterfront walkway.

\begin{table}[!t]
\caption{Weights for Corner-Weighted Model}
\label{tab:weights_corner_model}
\begin{tabularx}\columnwidth{l|Y}
\hline
\multicolumn{1}{c|}{\textbf{Cycle Segment}}    & \textbf{Weight} \\ \hline
Cawston (Road + Cycleway, Gordon to Water St.) & 0.005           \\ \hline
Ethel (Clement to Sutherland)                  & 0.95            \\ \hline
City Park (South of Yacht Club to Bridge)      & 2.25            \\ \hline
Waterfront (North of Yacht Club to Sunset Dr.) & 2.75            \\ \hline
All other roads, lanes and paths               & 1.0             \\ \hline
\end{tabularx}
\end{table}

In order to improve the relationship between the two data sets and achieve similar splits with the bike-share data as with the counter data, the \emph{Corner-weighted }model was developed and applied.  In this model, the weights provide an indication of how dissimilar the bike-share data is from the counter data at a given location.  By adjusting the weights, a similar split of traffic was achievable (Table~\ref{tab:weights_corner_model}).
Weights for the lakeside cycleway were made two to three times less favorable than any other road and Cawston 200 times  more favorable but this model still did not result in a strong model of the relationship between bike-share and counter data.  In practice, the demographics and location of bike-share riding behaviours is significantly different than the general riding population.

Based on the assessment of counter-split up-scaling the \emph{Corner-weighted}, \emph{Shortest Length} and \emph{Weighted Length} were selected as candidate models.

\subsection{Linear Regression of City Park bike-share Counts}

As the bike-share traffic pattern at City Park is similar to the traffic pattern recorded by the counters, the location was used to compare the relationship between counter data and bike-share counts resulting from each of the path-finding models.  A linear regression of weekly counter counts vs weekly bike-share counts was performed.   

Table~\ref{tab:linear_regression} compares the R-squared value and p-value for each of the path-finding models. The R-squared value and p-value are used to compare the regression models. The closer the R-squared is to one, the better. The p-value is an indication of whether the multiplier for the bike-share data is statistically significant. A p-value of less than 0.05 means that the multiplier is statistically significant. 

\begin{SCtable*}[][!t]
\caption{Linear regression of counter vs. bike-share data at City Park}
\label{tab:linear_regression}
\begin{tabular}{l|c|c}
\hline
\multicolumn{1}{c|}{\textbf{Model}} & \textbf{R-Squared Value} & \textbf{P-Value} \\ \hline
SL              & 0.3463 & 0.0343 \\ \hline
PP             & 0.1998 & 0.1260 \\ \hline
SPP   & 0.2125 & 0.1130 \\ \hline
WL            & 0.2341 & 0.0939 \\ \hline
CC         & 0.3128 & 0.0469 \\ \hline
UB                    & 0.2316 & 0.0959 \\ \hline
WC             & 0.2168 & 0.1089 \\ \hline
WCL       & 0.2376 & 0.0911 \\ \hline
\end{tabular}
\end{SCtable*}

The \emph{Shortest Length} and \emph{Weighted Corner} models were evaluated with the linear regression for all the counter locations to better understand the overall impact of weighting. The \emph{Weighted Corner} model performed better than \emph{Shortest Length} at Waterfront, but not for the other three locations.   It was determined that intentionally weighting the model to try to match the counter data did not result in creating a similar traffic pattern at each counter location.  As a result, the \emph{Shortest Length}, \emph{Closeness Centrality} and \emph{Weighted Corner Length} were selected as candidate models under this condition. 

\subsection{Determining Best Path-Finding Model}

\begin{figure*}[!t]
\captionsetup[subfigure]{justification=centering}
\begin{subfigure}[t]{0.49\textwidth}
	\centering
	\includegraphics[width=\textwidth]{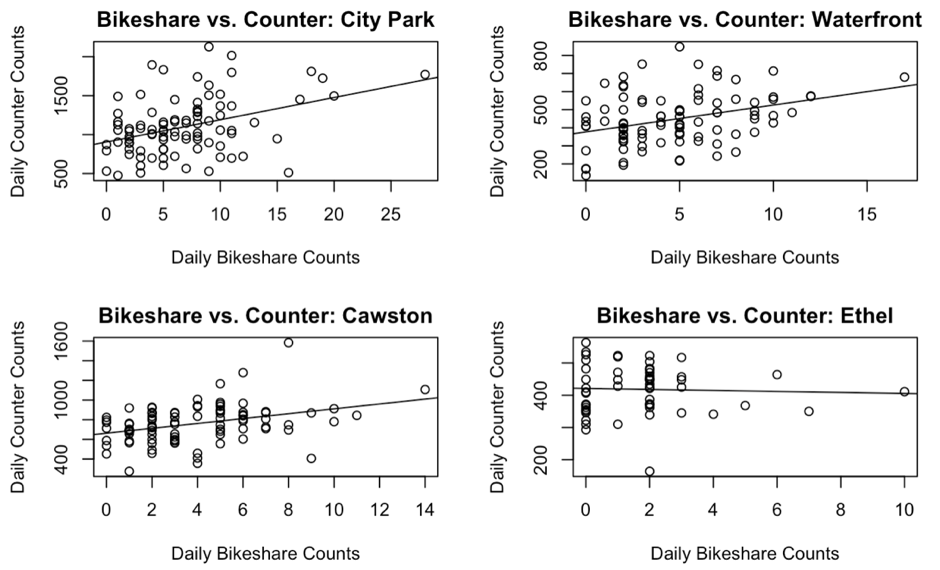}
	\caption{Relationship between daily bike-share vs counter data at each location}
	\label{fig:counter_vs_bike-share}
	\end{subfigure}
	\hfill
	\begin{subfigure}[t]{0.49\textwidth}
	\centering
	\includegraphics[width=\textwidth]{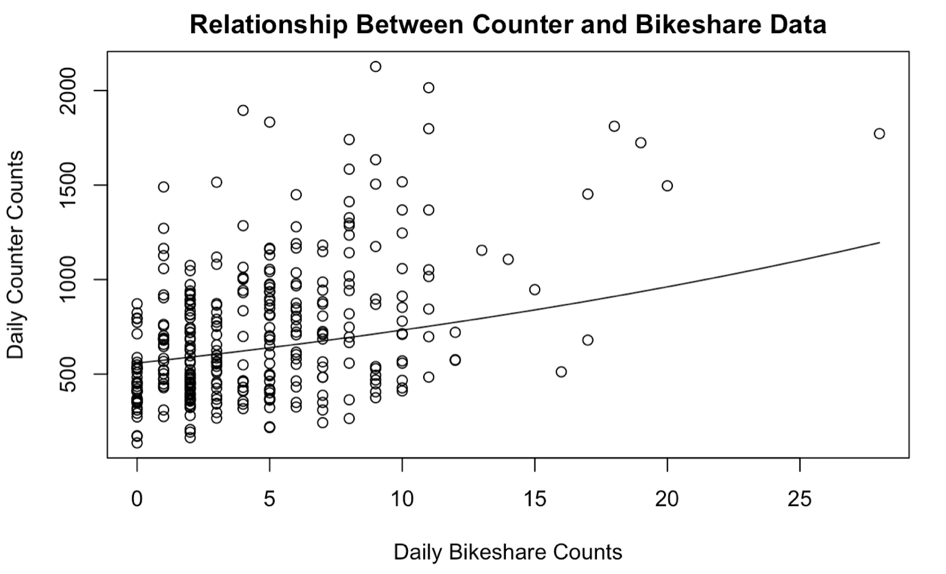}
	\caption{Mixed effects model between counter and bike-share data}
	\label{fig:mixed_model}
	\end{subfigure}
	\caption{Development of the mixed-effects model, for the relationship between counter and bike-share data}
\end{figure*}

\begin{SCtable*}[][!t]
\caption{Top ranking models based on path-finding strategies}
\label{tab:model_path_finding_summary}
\begin{tabular}{l|c|c|c|c}
\hline
\multicolumn{1}{c|}{\textbf{Model}} & \textbf{Visual} & \textbf{Speed} & \textbf{Split} & \textbf{\begin{tabular}[c]{@{}c@{}}Linear\\Regression\end{tabular}} \\ \hline
\multicolumn{1}{l|}{SL}            & 3 & 1 & 2 & 1 \\ \hline
\multicolumn{1}{l|}{PP}            & 1 &   &   &   \\ \hline
\multicolumn{1}{l|}{SPP} &   &   &   &   \\ \hline
\multicolumn{1}{l|}{WL}            &   &   & 3 &   \\ \hline
\multicolumn{1}{l|}{CC}       &   & 2 &   & 2 \\ \hline
\multicolumn{1}{l|}{UB}                   &   &   &   &   \\ \hline
\multicolumn{1}{l|}{WC}            & 2 & 3 & 1 &   \\ \hline
\multicolumn{1}{l|}{WCL}     &   &   &   & 3 \\ \hline
\end{tabular}
\end{SCtable*}

Table~\ref{tab:model_path_finding_summary}  summarizes the results from the evaluations of each path-finding model. The numbers indicate the order of the top 3 models for the particular evaluation.

Although Shortest Length presented minor issues in the visual evaluation, it was in the top two for all other evaluations. As it is based on the length of the streets, it was one of the more objective path-finding models. It was not weighted based on path preference or trying to fit the counter data. Therefore, this was the model selected to calculate AADB volumes.

\subsection{Calculating Average Daily Bicycle (AADB) Volume}

Calculating the AADB at each counter was done by finding the relationship between counter and bike-share data at each counter. To find the AADB for all downtown street segments, the effect each counter has on the AADB is required to be estimated.  Figure~\ref{fig:counter_vs_bike-share} demonstrates the relationship between counter and bike-share data and that the relationship is different for each counter.

\begin{figure}[!t]
\captionsetup[subfigure]{justification=centering}
\begin{subfigure}[t]{0.49\columnwidth}
\centering
	\includegraphics[width=\columnwidth]{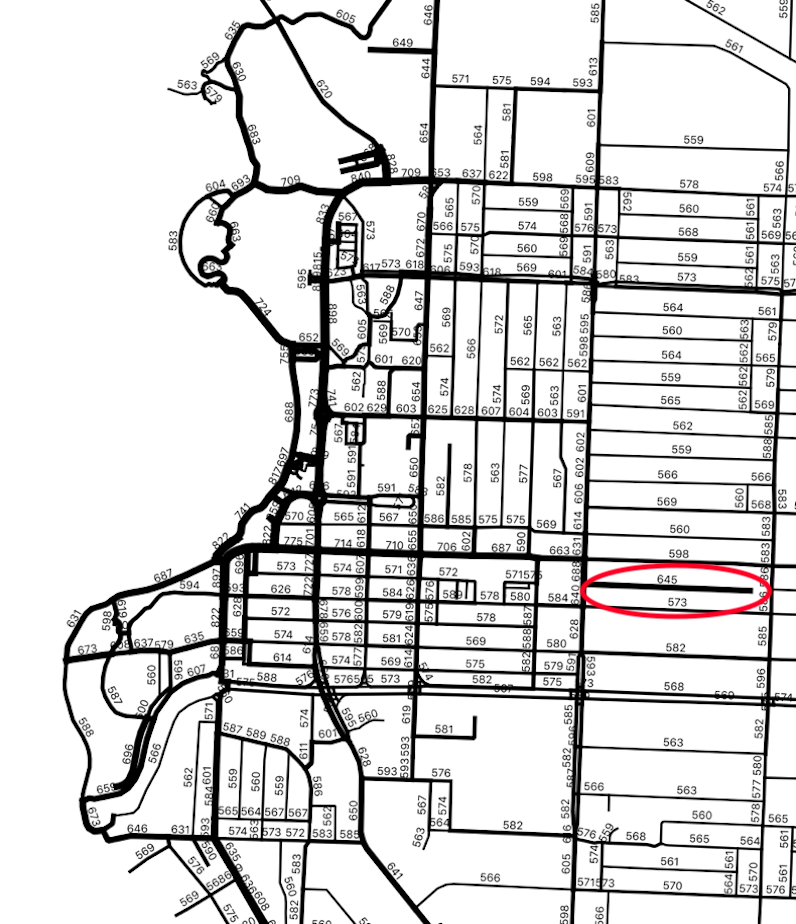}
	\caption{AADB for street segments based on the mixed effects model approach}
	\label{fig:AADB}
\end{subfigure}
\begin{subfigure}[t]{0.49\columnwidth}
	\centering
	\includegraphics[width=\columnwidth]{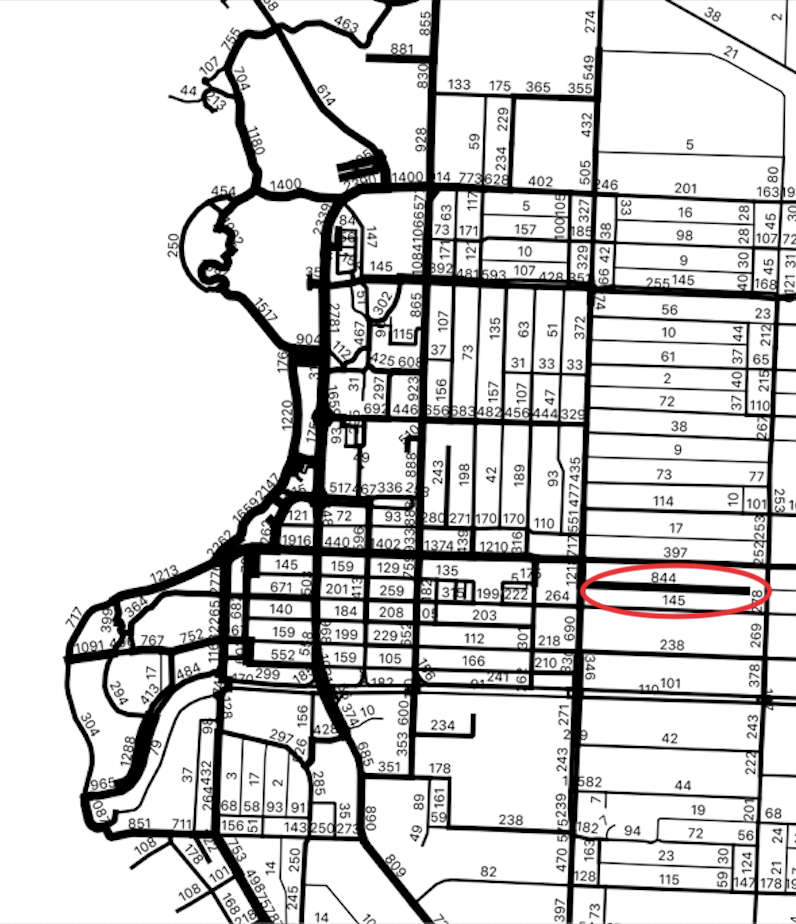}
	\caption{AADB for street segments calculated with least squares optimization approach}
	\label{fig:ls_AADB}
\end{subfigure}
\caption{A comparison of visualizations of annual average daily bike (AADB) volumes using the mixed effects and least-squares optimization models for paths in the cycling network}
\end{figure}

A mixed effects model is a generalized regression model that takes into account the effect of each counter location as it calculates the overall relationship between the counter and bike-share data. Since the count data includes multiple zeros and repeated values, the relationship was calculated using a negative binomial distribution.  Using this model, the overall general relationship between counter and bike-share data is shown in Figure~\ref{fig:mixed_model}.

Using the bike-share data captured during the study period, an average daily bike-share volume for bike-shares (AADB$_{bike-share}$) is represented as

\begin{eqnarray}
AADB_{bike-share} &=& \frac{\textrm{total bike-share volume}}{91\textrm{ days}}.\label{eqn:bs}
\end{eqnarray}
Using Equation~\ref{eqn:bs} and the relationship between counter and bike-share developed from the mixed effects model, a scale-up model to estimate the average daily bike (\emph{AADB}) was  determined to be
\begin{eqnarray}
AADB &=& e^{(0.02717094 \times AADB_{bike-share} + 6.325313)}  \label{eqn:AADB}.
\end{eqnarray}

Using Equation~\ref{eqn:AADB}, the AADB for each street segment was calculated and mapped in Figure~\ref{fig:AADB}. One of the drawbacks of finding one statistical model from the different counter models is that there is an averaging out effect. As a result, the same level of variance is not seen in the map of AADB values as is seen for the total bike-share counts. To retain the variation,  a single multiplier for the average daily bike-share data to approximately equal the AADB from the counter locations was determined.  The multiplier is determined through a least squares optimization, with the objective function:

\begin{eqnarray}
f(x) &=& \sum_{i=1}^{4} (ax_i - y_i)^2 w_i. \label{eqn:ls}
\end{eqnarray}
where for each location $i$, $a$ is the location multiplier, $x_i$ is the bike-share counts, $y_i$ is the counter counts, and $w_i$ is the split of counts at each counter location.

After minimizing Equation~\ref{eqn:ls}, the equation for AADB was determined to be
\begin{eqnarray}
AADB &=& 159 \times \frac{\textrm{total bike-share count}}{91}.\label{eqn:final_AADB}
\end{eqnarray}

Figure~\ref{fig:ls_AADB} shows the AADB for each street.  This technique estimates the AADB without losing the variation as compared to the mixed effects model.

\subsection{Insights}

\subsubsection{Infrastructure Gaps}
From the analysis, it was determined that  Bernard Ave. and Pandosy St./Water St. were highly travelled by bike-share users, even though there are no bike lanes for cyclists. Bike-share users either cycled with traffic or on the sidewalks. There are also gaps in infrastructure at Clement and Doyle, which were often cycled by bike-share users.

\subsubsection{Highway Crossings}
Table~\ref{tab:hwy97_crossing} summarizes the number of bike-share trips that crossed Highway 97 at each intersection and the tunnel at City Park.  As previously mentioned, the \emph{Shortest Path} algorithm had trouble finding paths at the Highway 97 and Abbott intersection due to GPS inaccuracy. 

\begin{SCtable}[][!t]
\caption{Total number of bike-share trips at each crossing of Highway 97}
\label{tab:hwy97_crossing}
\begin{tabular}{l|c}
\hline
\multicolumn{1}{c|}{\textbf{\begin{tabular}[c]{@{}c@{}} Highway\\Crossing Point\end{tabular}}} & \textbf{ \begin{tabular}[c]{@{}c@{}}Number of\\bike-share Trips\end{tabular}}                                          \\ \hline
Tunnel at City Park & 622 \\
Abbott              & 426 \\
Pandosy/Water                                        & 504 (south), 207 (north)\\
Ellis               & 312 \\
Richter                                              & 165 (south), 216 north)\\
Ethel               & 164 \\ \hline
\end{tabular}
\end{SCtable}

\subsubsection{Use of Laneways}
In the analysis of AADB values for city segments, one laneway stood out as being exceptionally popular.  The laneway is located in a residential area and is circled in Figure~\ref{fig:AADB} and~\ref{fig:ls_AADB}.

The laneway is parallel to and directly south of Bernard Avenue, a major roadway to/from downtown without significant infrastructure for cycling. It indirectly connects Richter to Ethel through a park at the east end of the laneway. Richter has a bike lane and Ethel has a separated, paved path.  The route through the park is clearly shown on Google Maps and provides a bypass for significantly more congested and dangerous roadways.  Additionally, a bike-share haven at the southeast corner of the park, where bike-share users could initiate or end a trip provided a focal point for traffic in and out of the downtown core. 

Outside of this one laneway of note, the laneways in the residential blocks are used less than the residential streets.  While laneways can provide safer cycling routes, residential laneways often do not directly connect two streets or cyclists may be treating those laneways as private property. In the commercial blocks, the laneways are used as much as the tertiary downtown streets, since they connect to streets on both sides and provide safe bypass options for heavily congested and busy streets. 

\section{Conclusions and Future Work}\label{sec:conclusion}

Using a combination of bike-share and counter data, an analysis of the City of Kelowna's bicycle network was conducted.  Using OSMnx allowed for the mapping of lower accuracy, infrequent GPS points to valid intersections, finding feasible cycling paths to connect those GPS points for each trip, and analysis of the structure of the bike network. Based on the four evaluation criteria, the best path-finding model was \emph{Shortest Length}, which weighted street segments based on their length.

In terms of bike-share traffic patterns, it was determined that for the region, there were more recreational bicycle traffic recorded by the counters at City Park and Waterfront. The bike-share traffic pattern is different from the traffic recorded by the counters at Cawston and Ethel which represents areas of high commuter use.  Further, the mixed effects model and least squares optimization provided estimates of AADB within the cycling network. 

While the mixed effects model and least squares optimization provided estimates of AADB, they are only estimating how the bike-share data scaled up to overall volumes. As bike-share traffic patterns are significantly different from the general traffic, conclusions can only be drawn as to how the bike network was used by bike-share riders with  acknowledgement to the different riding behaviours between groups. 

Key insights were uncovered regarding infrastructure gaps and informal biking routes.  Some major streets were highly travelled by bike-share users, even though they lack infrastructure for cyclists or have infrastructure gap providing insight for future bike route planning.  Further, additional insights were gained into highway crossings and laneway usage to bypass areas of congestion and infrastructure gaps.  In future work, more data is required based on GPS positioning for representative cyclists to further improve and validate models.  Additionally, deploying additional counters within the network will provide more data on how all cyclists use the network.

\section{Acknowledgment}

The authors gratefully acknowledge the support of Matt Worona and Kamil Rogowski from the City of Kelowna for supporting this research.

\bibliographystyle{spmpsci}
\bibliography{refs}

\end{document}